\documentclass[trackchanges,twocolumn]{aastex701}

\usepackage{amsmath}

\begin{document}

\title{Extracting redshifts from 2D slitless spectroscopic images using deep learning for the CSST galaxy survey}

\author[0000-0001-7283-1100, sname='Xingchen']{Xingchen Zhou}
\affiliation{National Astronomical Observatories, Chinese Academy of Sciences, 20A Datun Road, Beijing 100101, People's Republic of China}
\affiliation{Science Center for China Space Station Telescope, National Astronomical Observatories, Chinese Academy of Sciences, 20A Datun Road, Beijing 100101, People's Republic of China}
\email{xczhou@nao.cas.cn}

\author[0000-0003-0709-0101, sname='Yan']{Yan Gong}
\affiliation{University of Chinese Academy of Sciences, Beijing, 100049, People's Republic of China}
\affiliation{National Astronomical Observatories, Chinese Academy of Sciences, 20A Datun Road, Beijing 100101, People's Republic of China}
\affiliation{Science Center for China Space Station Telescope, National Astronomical Observatories, Chinese Academy of Sciences, 20A Datun Road, Beijing 100101, People's Republic of China}
\email[show]{gongyan@bao.ac.cn}

\author[0000-0001-7314-4169, sname='Xin']{Xin Zhang}
\affiliation{National Astronomical Observatories, Chinese Academy of Sciences, 20A Datun Road, Beijing 100101, People's Republic of China}
\affiliation{Science Center for China Space Station Telescope, National Astronomical Observatories, Chinese Academy of Sciences, 20A Datun Road, Beijing 100101, People's Republic of China}
\email{zhangx@bao.ac.cn}

\author[sname='Xian-Min']{Xian-Min Meng}
\affiliation{National Astronomical Observatories, Chinese Academy of Sciences, 20A Datun Road, Beijing 100101, People's Republic of China}
\affiliation{Science Center for China Space Station Telescope, National Astronomical Observatories, Chinese Academy of Sciences, 20A Datun Road, Beijing 100101, People's Republic of China}
\email{mengxm@nao.cas.cn}

\author[0000-0003-0850-3641, sname='Haitao']{Haitao Miao}
\affiliation{National Astronomical Observatories, Chinese Academy of Sciences, 20A Datun Road, Beijing 100101, People's Republic of China}
\affiliation{State Key Laboratory of Radio Astronomy and Technology, National Astronomical Observatories, Chinese Academy of Sciences, Beijing 100101, People's Republic of China}
\email{miaoht@bao.ac.cn}

\author[0000-0002-8705-6327, sname='Run']{Run Wen}
\affiliation{State Key Laboratory of Dark Matter Physics, Tsung-Dao Lee Institute, Shanghai Jiao Tong University, Shanghai 201210, People’s Republic of China}
\affiliation{Department of Astronomy, School of Physics and Astronomy, Shanghai Jiao Tong University, Shanghai 200240, People’s Republic of China}
\email{wenrun1214@sjtu.edu.cn}

\author[0000-0001-6800-7389, sname='Nan']{Nan Li}
\affiliation{National Astronomical Observatories, Chinese Academy of Sciences, 20A Datun Road, Beijing 100101, People's Republic of China}
\affiliation{Science Center for China Space Station Telescope, National Astronomical Observatories, Chinese Academy of Sciences, 20A Datun Road, Beijing 100101, People's Republic of China}
\email{nan.li@nao.cas.cn}

\begin{abstract}
  Wide-field slitless spectroscopic galaxy surveys, such as the one performed by the upcoming Chinese Space Station Survey Telescope (CSST), are crucial for precision cosmology but present formidable data analysis challenges. Because spectra are dispersed directly onto the detector, they are convolved with the 2-dimensional (2D) spatial morphology, which complicates wavelength calibration and consequently degrades the fidelity of subsequent 1-dimensional (1D) spectral extraction. To overcome these limitations, we present a deep learning framework that extracts redshifts directly from 2D slitless spectral images, bypassing 1D extraction entirely. We construct a realistic mock dataset for the CSST $GV$ and $GI$ band using high-resolution images from HSC-SSP PDR3 and spectral energy distributions (SEDs) from DESI DR1. A Bayesian convolutional neural network implemented by Monte Carlo dropout is employed to map the 2D spectral images to redshift estimations while simultaneously quantifying uncertainties. We find that our model can achieve a precision $\sigma_{\rm NMAD}=0.0104$ and mean uncertainty $\langle E / (1 + z_{{\rm true}}) \rangle=0.0155$ for sources with ${\rm SNR}_{GI}\geq1$. For sources with ${\rm SNR}_{GI}$ higher than 3.0, 5.0 and 10.0, $\sigma_{\rm NMAD}$ can achieve 0.0047, 0.0037 and 0.0024 respectively, matching the redshift precision requirements for studies such as BAO using the CSST slitless spectroscopic surveys. Furthermore, by utilizing spatial augmentations, the network demonstrates resilience to wavelength calibration errors. This work provides a novel and robust pathway for data analysis of next-generation slitless spectroscopic galaxy surveys. 
  
  
\end{abstract}


\keywords{\uat{Space telescopes}{1547} -- \uat{Spectroscopy}{1558} -- \uat{Redshift Surveys}{1378} -- \uat{Convolutional neural networks}{1938} -- \uat{Bayesian Statistics}{1900}}


\section{Introduction} \label{sec:introduction}
Mapping the large-scale structure (LSS) of the Universe and tracing the evolution of galaxies across cosmic time fundamentally relies on acquiring accurate redshift measurements for vast number of extragalactic sources~\citep{Weinberg2013,Madau2014}. While photometric surveys can rapidly derive redshifts (photo-$z$) for billions of galaxies, these measurements inherently suffer from broad uncertainties and catastrophic outliers due to the low resolution of broadband filters~\citep{Salvato2019,Newman2022}. In contrast, spectroscopic observations capture the detailed emission and absorption features of galactic spectra, yielding accurate spectroscopic redshifts for high-precision cosmological constraints by Baryon Acoustic Oscillations (BAO)~\citep{Bassett2010,Ferreira2025} and Redshift-Space Distortions (RSD)~\citep{Hamilton1998}, as well as detailed characterization of galaxy kinematics and stellar populations~\citep{Conroy2013, Gallazzi2005, Cappellari2016}. 

Despite high spec-$z$ precision, traditional multi-object spectroscopic surveys, which utilize fibers or slit masks, are fundamentally bottlenecked by target pre-selection and the mechanical limitations of placing fibers in dense focal planes~\citep{Blanton2003,Aghamousa2016}. To overcome these challenges, space-base slitless spectroscopy has emerged as a powerful solution. By using slitless instruments, they can perform wide-field and highly multiplexed spectroscopic observations~\citep{Weiner2012,Brammer2012, Laureijs2011}. Ongoing and upcoming space-based missions, e.g. Euclid~\citep{Amendola2013}, Nancy Grace Roman Space Telescope~\citep{Mosby2020}, Chinese Space Station Survey Telescope (CSST, ~\citet{Gong2019,Gong2025,Gong2026}), utilize slitless spectroscopic surveys to capture the spectra of the luminous sources within their expansive fields of view, completely free from atmospheric interference. 

However, the rich data product of slitless spectroscopy introduces great challenges in data analysis. Because there is no slit to disperse light with high spectral resolution, the spectrum recorded on the detectors is 2D morphological shape convolved with point spread function (PSF). In densely populated survey fields, this leads to severe spectral overlapping and cross-contamination among neighboring objects. Furthermore, the slitless spectroscopy introduces severe complications for accurate wavelength calibration. Without a fixed spatial reference, the zero-point of the wavelength for each dispersed spectrum depends fundamentally on the exact spatial location and morphological centroid of the source. Any asymmetry in the galaxy or uncertainty in its position directly propagates into systematic wavelength shifts. Traditional data reduction pipelines attempt to calibrate these shifts, deblend overlaps, and extract 1D spectra from the 2D detector images, but this extraction process is highly complex, prone to compounding systematic errors, and inevitably results in a significant loss of spectral information~\citep{Brammer2012, Kummel2009, Momcheva2016}. 

For CSST, these challenges are more severe. Unlike other slitless spectroscopic surveys that typically acquire direct, undispersed images of the target field alongside the dispersed observations to pinpoint source coordinates, the specific focal plane design of CSST excludes simultaneous direct imaging during survey~\citep{Gong2026}. Consequently, establishing the astrometric centroid of a source and the zero-point required to anchor the wavelength, necessitates cross-matching the dispersed spectra with separately and independently acquired photometric catalogs. Any inherent astrometric uncertainties, such as focal plane distortions and alignment errors between these distinct observations propagate directly into systematic shifts in the assigned wavelengths, severely affecting the 1D extraction pipelines. On the other hand, the issue of spectral overlapping can be safely neglected, since the estimated source density is only $\sim2-3$ galaxies per $\rm arcmin^2$~\citep{Gong2019}.

Deep learning has rapidly emerged as a powerful tool to overcome these data analysis challenges. In particular, convolutional neural networks (CNNs) are exceptionally well-suited for astronomical imaging due to their ability to learn hierarchical and non-linear feature representations directly from complex and high-dimensional data~\citep{Lecun2015,Dieleman2015,Hezaveh2017}. Unlike traditional parametric modeling, which relies on rigid assumptions and hand-crafted features, deep learning models can automatically identify optimal mappings between noisy observational inputs and physical parameters, making them highly resilient to complex systematics and survey-specific artifacts. Leveraging deep learning allows us to fundamentally rethink how slitless spectra are analyzed: by operating directly on the 2D spectral images. This approach bypasses the complicated and error-prone 1D extraction step entirely. By treating the dispersed 2D image as the primary data product, CNNs can jointly utilize both the spectral dispersion features and the spatial morphological information, maximizing the outcomes from low signal-to-noise ratio (SNR) observations.

In this work, we present a deep learning framework to estimate redshifts directly from 2D slitless spectral images, tailored specifically for the upcoming CSST spectroscopic survey. Due to the focal plane design of CSST, we concentrate on wavelength calibration. To train and validate our models, we construct a highly realistic mock dataset of 2D spectral images, combining high-resolution imaging from the Hyper Suprime-Cam (HSC,~\citet{Aihara2018, Aihara2022}) with precise spectral templates from the Dark Energy Spectroscopic Instrument (DESI,~\citet{Abareshi2022,Karim2025}). Recognizing the critical importance of error characterization in precision cosmology, we employ the Bayesian neural network (BNN,~\citet{Mullachery2018,Goan2020}). This network allows our model not only predict accurate point estimates for the spec-$z$s, but also provide robust and statistically reasonable epistemic and aleatoric uncertainties for every measurement. 

This paper is organized as follows: Section~\ref{sec:datasets} describes the generation of the mock slitless spectroscopic dataset. In Section~\ref{sec:methodology}, we discuss the details of the neural network architectures employed in this study. We present the performance of our framework in Section~\ref{sec:results}, followed by several discussions in Section~\ref{sec:discussions}. Finally, we summarize our results in Section~\ref{sec:conclusion}.

\section{Datasets} \label{sec:datasets}
We simulate the 2D slitless spectra using \texttt{sls\_1d\_spec}~\footnote{\url{https://csst-tb.bao.ac.cn/code/zhangxin/sls\_1d\_spec}}, a core software module within the CSST simulation pipeline. Unlike the earlier version detailed in~\citet{Zhou2024accurately}, this updated version introduces the critical capability to input high-resolution and ideal images directly into the spectral generation process. This advancement enables the incorporation of highly realistic and complex galaxy morphologies across the three CSST spectroscopic bands, i.e. $GU, GV$ and $GI$, thereby overcoming the limitations of relying on simply assumed parametric S\'ersic profiles. To ensure our mock dataset is as realistic as possible, we utilize observational data for the simulations. Specifically, we extract high-fidelity spatial morphologies from the Hyper Suprime-Cam Subaru Strategic Program Public Data Release 3 (HSC-SSP PDR3,~\citet{Aihara2022}) and pair them with corresponding Spectral Energy Distributions (SEDs) retrieved from the Dark Energy Spectroscopic Instrument Data Release 1 (DESI DR1,~\citet{Karim2025}).

We select sources from DESI DR1 catalog based on the following criteria:
\begin{equation}\label{eq:desi}
  \begin{aligned}
    &\rm \texttt{ZCAT\_PRIMARY == True}\\
    &\rm \texttt{MASKBITS == 0}\\
    &\rm \texttt{SPECTYPE == GALAXY}\\
    &\rm \texttt{ZWARN == 0}\\
    &\rm \texttt{FLUX\_G, R, Z > 0} \\
    &\rm \texttt{FLUX\_IVAR\_G, R, Z > 0} \\
    &\rm \texttt{SHAPE\_R < 1.0}
  \end{aligned}
\end{equation}
The \texttt{ZCAT\_PRIMARY} flag ensures the selection of the optimal recommended redshift in cases of duplicate observations, while \texttt{MASKBITS == 0} guarantees that the source does not overlap with any masked regions. The \texttt{SPECTYPE} and \texttt{ZWARN} parameters correspond to the source classification and redshift fitting reliability flags derived by the redshift measurement software, \texttt{Redrock}~\footnote{\url{https://github.com/desihub/redrock}}, respectively. We strictly require a confirmed galaxy classification with high-quality redshift measurement. To further ensure photometric quality, we select sources with positive fluxes and inverse variances across the $g, r$ and $z$ bands, as measured by the \texttt{Tractor} algorithm~\footnote{\url{https://thetractor.org/}}. Furthermore, we restrict the half-light radius, \texttt{SHAPE\_R} to be less than 1.0 arcsec. This morphological constraint is specifically applied to increase the signal-to-noise ratio (SNR) of the resulting mock spectra. Because the dispersed light at each wavelength is spatially convolved with the 2D morphology of the source, a more compact radius concentrates the flux over fewer detector pixels. This effectively boosts the signal relative to the background and instrumental noise, thereby enhancing the overall SNR. Sources with a measured radius of zero, which indicates an unresolved, PSF morphology, are intentionally preserved in our sample, provided that they are still confidently classified as galaxies during the redshift fitting process.  

Sources from the HSC-SSP PDR3 are selected in wide Spring field based on the following criteria:
\begin{equation}\label{eq:hsc}
  \begin{aligned}
    &\rm \texttt{i\_cmodel\_mag < 24.0} \\
    &\rm \texttt{i\_cmodel\_flux > 0} \\
    &\rm \texttt{i\_cmodel\_fluxerr > 0} \\
    &\rm \texttt{i\_cmodel\_flux / i\_cmodel\_fluxerr > 100}
  \end{aligned}
\end{equation}
For simplicity, we restrict our 2D morphological templates to HSC $i$-band images, which demonstrates highest SNR among all bands for considered sources. The upper magnitude limit is set to 24.0, sufficiently encompassing the anticipated depth of the CSST wide-field slitless spectroscopic observations, i.e. 23.2 in $GU$, 23.4 in $GV$ and 23.2 in $GI$ respectively~\citep{Gong2025}. Additionally, we enforce a strict SNR threshold exceeding 100 at $i$ band. This stringent constraint is crucial, since it guarantees that the selected photometric images possess sufficient fidelity to be treated as ideal and noise-free templates. As a result, these high-SNR images can be directly utilized in the mock spectra generation pipeline without further noise calibration to match the expected CSST instrumental noise levels. 

Subsequently, we cross-match the DESI and HSC catalogs using a maximum matching radius of 1.5 arcsec, yielding a sample of approximately 700,000 galaxies. Image cutouts and their corresponding PSFs are retrieved via the PDR3 data access service~\footnote{\url{https://hsc-release.mtk.nao.ac.jp/doc/index.php/tools/}}. For each cutout, the central target galaxy is identified and isolated using the \texttt{photutils}~\footnote{\url{https://photutils.readthedocs.io/en/stable/index.html\#}} package. To recover the intrinsic morphological structure, the isolated galaxy is deconvolved with its corresponding PSF using the Richardson-Lucy deconvolution algorithm~\citep{Richardson1972}, implemented in \texttt{scikit-image}~\footnote{\url{https://scikit-image.org/}}, thereby the ideal galaxy images are produced. In parallel, we construct the SEDs from model spectra for each source using the linear combination of the DESI basis templates and their corresponding coefficients. Finally, these ideal images and their paired SEDs are fed into the simulation software to generate the mock 2D slitless spectra. 

Note that we restrict our simulations to the $GV$ and $GI$ bands. This choice is necessitated by the wavelength coverage of the DESI instrument, which truncates at approximately $4000\rm \AA$ at the blue end~\citep{Abareshi2022,Aghamousa2016}, insufficient to cover the CSST $GU$ bandpass. Furthermore, this restriction is consistent with practical observational expectations, as the anticipated SNR in the $GU$ band is generally low to reliably estimate redshifts. Following this comprehensive procedure, our final dataset comprises 689,295 mock 2D slitless spectral images along with their 1D extracted spectra. 

\begin{figure}[h]
  \centering
  \includegraphics[width=1\linewidth]{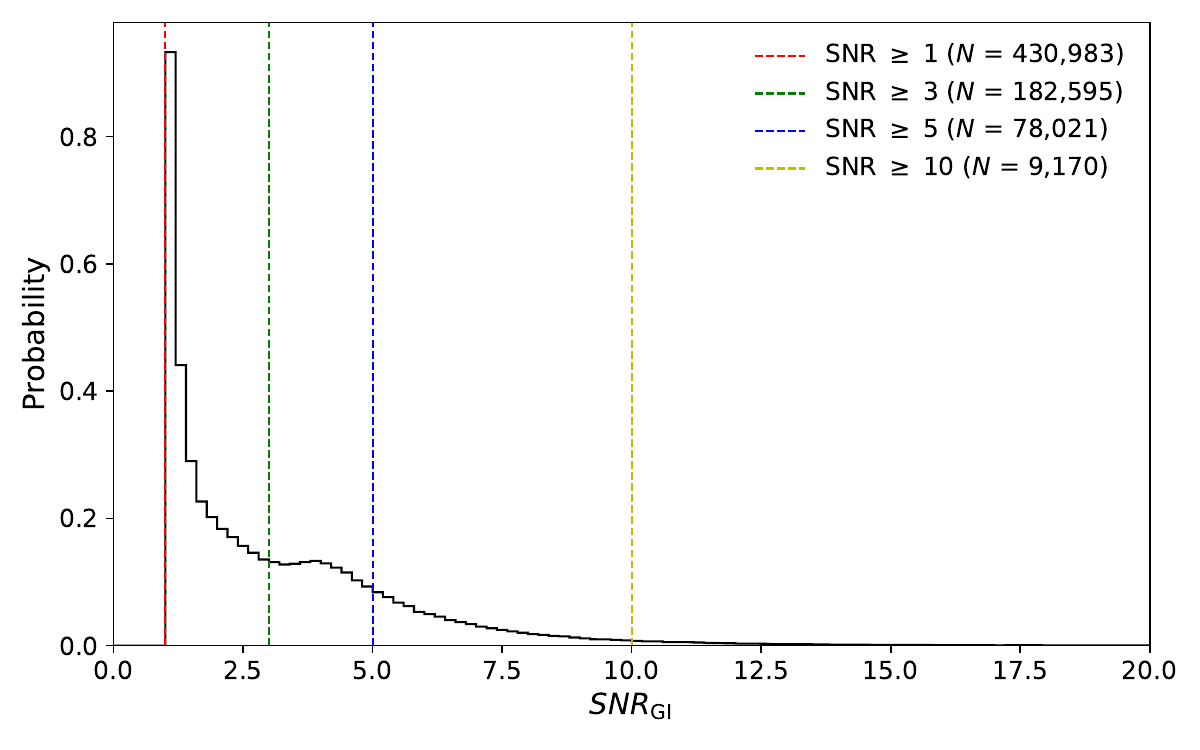}
  \caption{The distribution of SNR at $GI$ band, ${\rm SNR}_{GI}$. The red, green, blue and yellow lines indicate the SNR thresholds of 1, 3, 5 and 10, respectively. The number of sources exceeding each threshold is provided in the legend.}
  \label{fig:snr_dist}

\end{figure}

\begin{figure}
    \centering
    \includegraphics[width=1\linewidth]{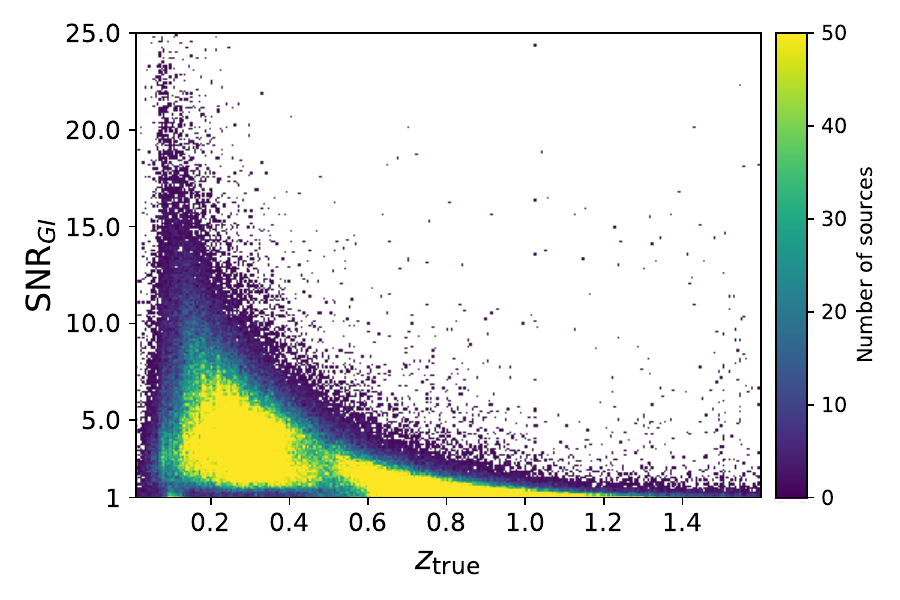}
    \caption{The density plot of ${\rm SNR}_{GI}$ with respect to $z_{\rm true}$. The color bar indicates the number of sources within each pixel. }
    \label{fig:snr_vs_z}
\end{figure}

In this work, we restrict our analysis to sources with a mean SNR calculated by 1D spectra at the $GI$ band with ${\rm SNR}_{GI} \geq 1$. The mean SNR is defined as 
\begin{equation}\label{eq:snr}
  {\rm SNR} = \frac{1}{n}\sum_i{\frac{\rm f_i}{\rm e_i}}
\end{equation}
where $f_i$ and $e_i$ denote the flux and its associated error at wavelength $i$, respectively. During this calculation, any negative flux values are artificially set to 0. The distribution of ${\rm SNR}_{GI}$ for the samples is presented in Figure~\ref{fig:snr_dist}. To characterize the data quality, we compute the number of sources exceeding specific SNR thresholds: enforcing ${\rm SNR}_{GI} >$ 1, 3, 5 and 10 yields sample sizes of 430,983, 182,596, 78,021 and 9,170, respectively. Furthermore, the density plot of ${\rm SNR}_{GI}$ versus $z_{\rm true}$ is presented in Figure~\ref{fig:snr_vs_z}, illustrating a clear decrease in the SNR as redshift increases.

\begin{figure*}
  \centering
  \includegraphics[width=0.45\linewidth]{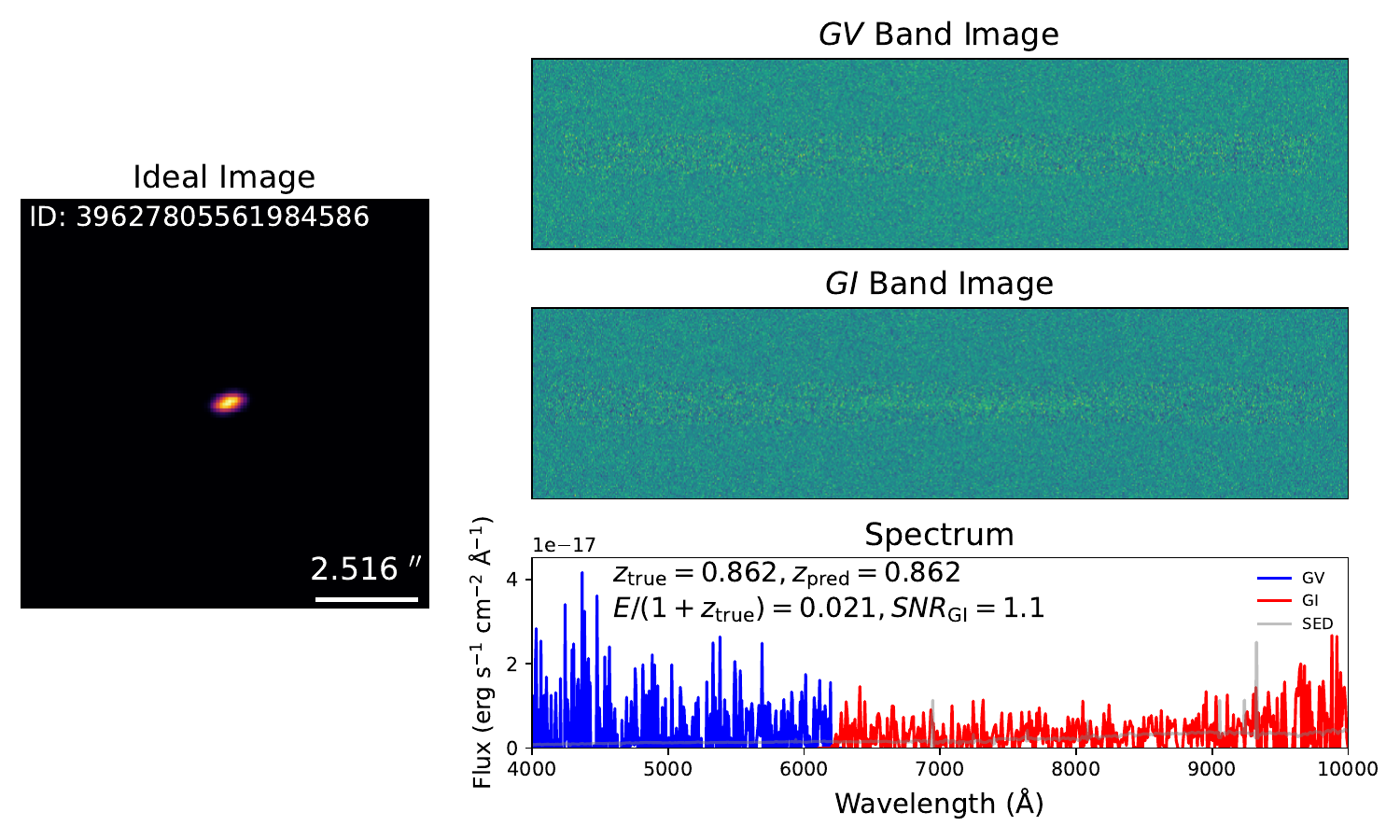}
  \includegraphics[width=0.45\linewidth]{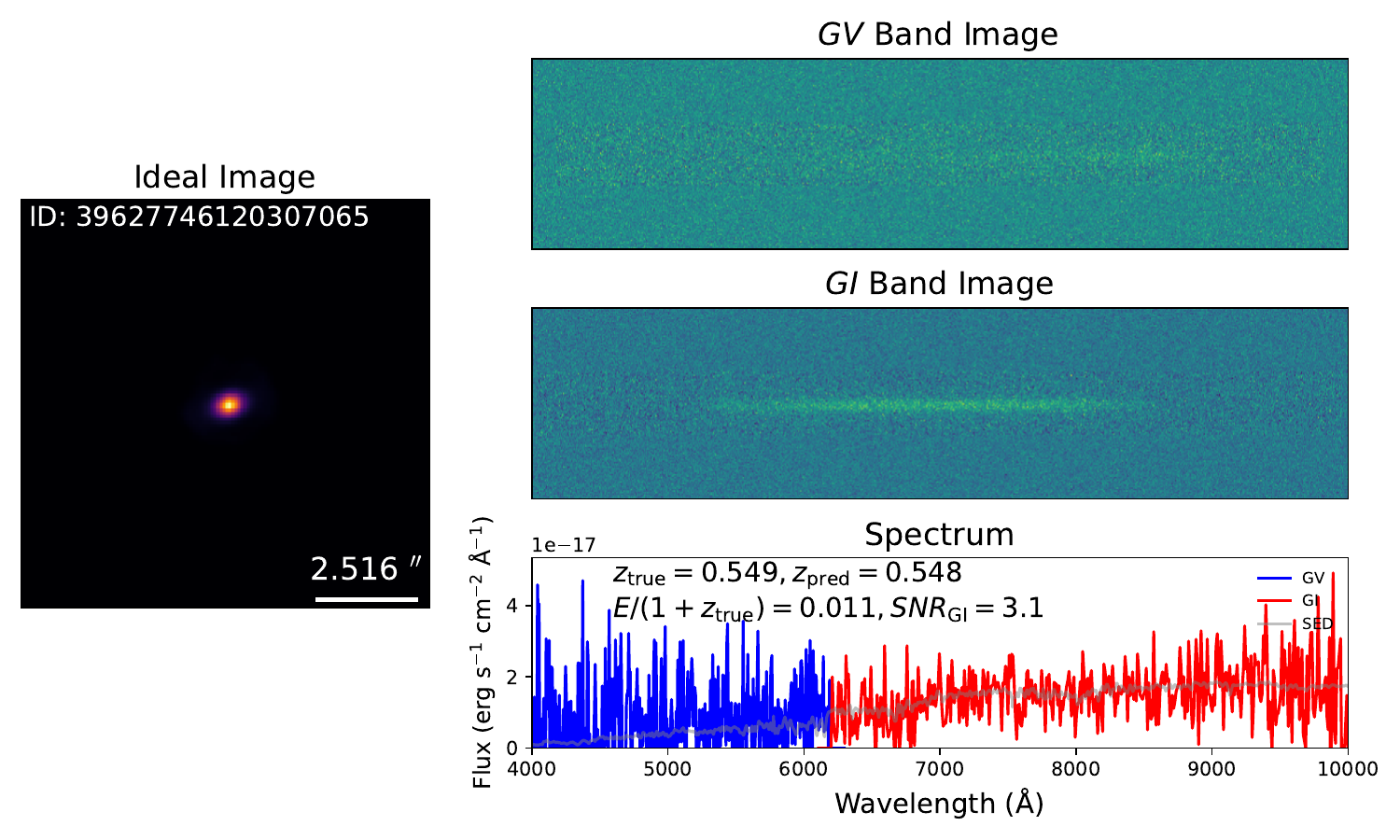}
  \\
  \includegraphics[width=0.45\linewidth]{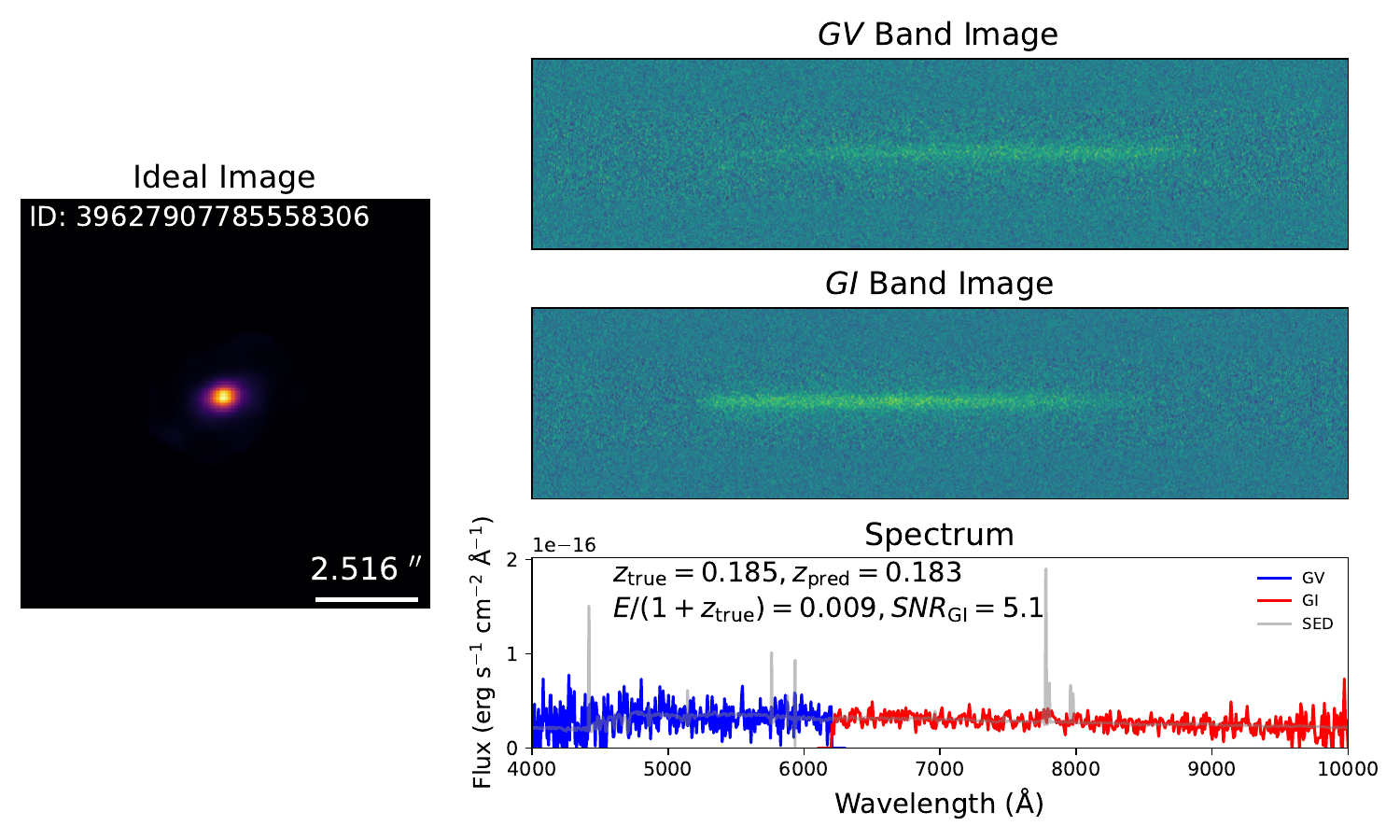}
  \includegraphics[width=0.45\linewidth]{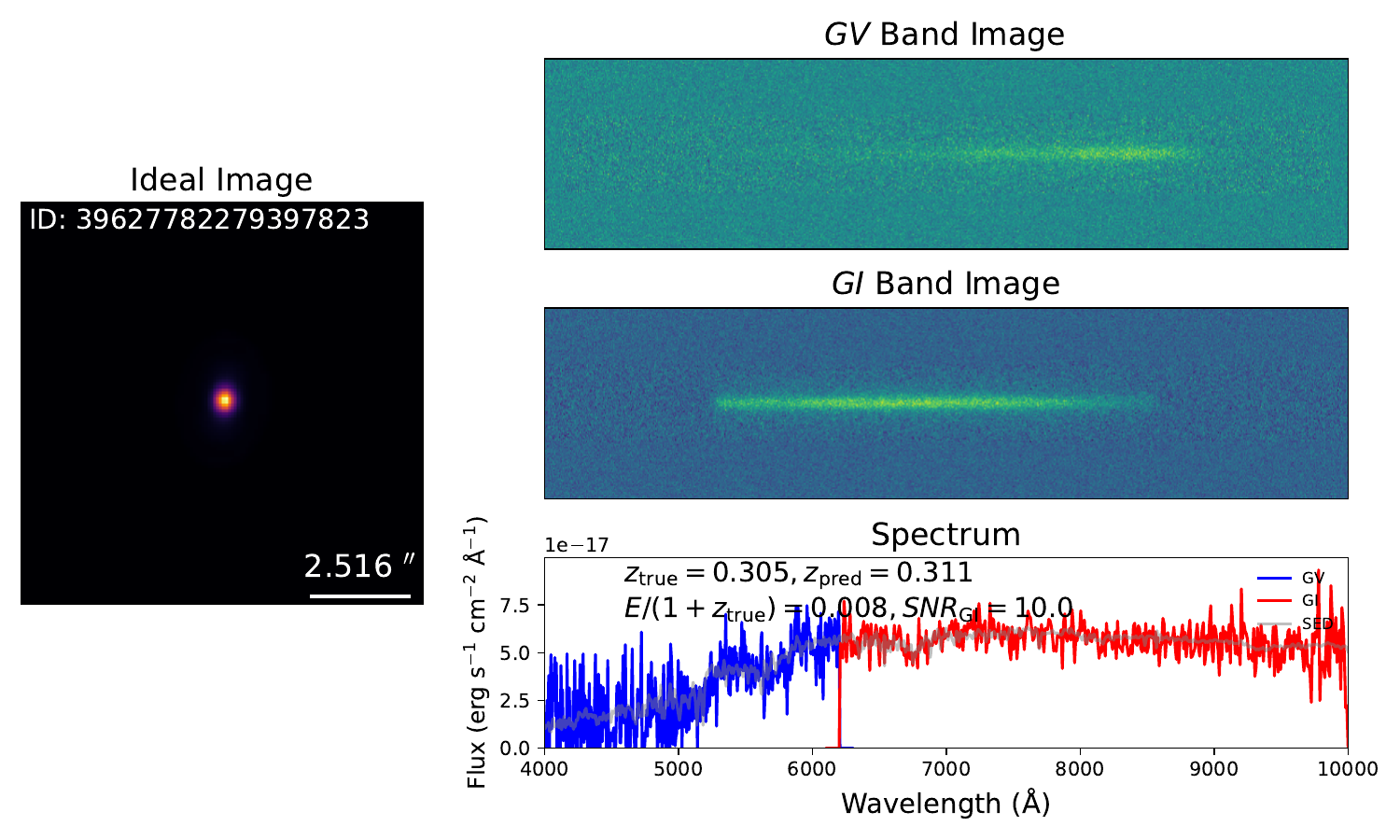}
  \caption{Four representative examples of mock slitless spectra at various redshifts and ${\rm SNR}_{GI}$. For each example, the leftmost panel displays the ideal spatial image, while the two upper right panels illustrate the corresponding 2D spectral images in the $GV$ and $GI$ bands. The lower panel compares the intrinsic SED with the extracted 1D spectrum.}
  \label{fig:data_example}
\end{figure*}

Representative examples of the simulated 2D spectral images, alongside their corresponding 1D spectra and ideal image cutouts, are illustrated in Figure~\ref{fig:data_example}. Each ideal image panel is annotated with their DESI target ID. Furthermore, the 1D spectrum panels display the mean ${\rm SNR}_{GI}$, the true spectroscopic redshift $z_{\rm true}$, the model's predicted redshift $z_{\rm pred}$, and the associated normalized uncertainty $E/(1 + z_{\rm true})$ as defined in Section~\ref{sec:results}. 

\section{Methodology} \label{sec:methodology}
In this section, we introduce the employed Bayesian convolutional neural network and the training procedure. 
\subsection{Bayesian neural network} \label{sec:bayesian neural network}

\begin{table}
\caption{The architecture of employed neural networks.}
\label{tab:arch}
\begin{tabular}{ccc}
\hline
Layer     & Output Shape & Param      \\ \hline\hline
Input     & (2, 40, 480) & -          \\ \hline
ResNet-34 & 512          & 21,281,536 \\ \hline
FC        & 256          & 131,328    \\ \hline
GELU      & 256          & -          \\ \hline
Dropout   & 256          & -          \\ \hline
FC        & 128          & 32,896     \\ \hline
GELU      & 128          & -          \\ \hline
Dropout   & 128          & -          \\ \hline
Output    & 1 / 2\ $^a$        & 129 / 258  \\ \hline
\end{tabular}
\\ \\
$^a$ The deterministic and Bayesian neural networks have 1 and 2 output values respectively. 
\end{table}

The 2D spectral images from the $GV$ and $GI$ bands in array shape of (40, 480) are stacked along the channel dimension, forming a tensor of shape (2, 40, 480). To estimate redshifts directly from these stacked images, we employ a Bayesian Convolutional Neural Network (BCNN). The network architecture comprises two primary components: a CNN feature extractor and a multi-layer perceptron (MLP) estimator. The feature extractor is based on the ResNet-34 architecture~\citep{He2015deep}, with the input layer modified to accommodate our specific data dimensions. This component is designed to compress the complex 2D spatial and spectral information into a lower-dimensional latent representation. The MLP estimator subsequently maps these latent vectors to the final redshift predictions. The estimator comprises two fully connected (FC) layers, each followed by a GELU activation function~\citep{Hendrycks2016} to introduce non-linearity. Additionally, a dropout layer with dropout rate of 0.5 is appended after each activation. The network architectures are demonstrated in Table~\ref{tab:arch}. 

In standard deterministic networks, dropout is only activated during the training phase. However, for the Bayesian network, the dropout is kept active during both the training and inference stages to enable the Monte Carlo (MC) Dropout method~\citep{Gal2015dropout}. By performing multiple forward passes for a single input during testing, this approach effectively approximates a Bayesian ensemble, allowing robust quantification for the epistemic (model-dependent) uncertainty. 

To simultaneously capture the aleatoric (data-dependent) uncertainty inherent to the noisy spectral observations, we design the final layer of the network to output two distinct values: the mean redshift prediction and the log-variance. Predicting the variance in logarithmic scale ensures that the recovered variance remains strictly positive. Together, these two outputs parameterize a Gaussian posterior distribution for each source. While more complex distributions could theoretically be employed, a Gaussian profile serves as a robust and physically reasonable way for modeling redshift uncertainties in this context. 

The total uncertainties for a given source is derived by combining both the aleatoric and epistemic components obtained from the MC dropout sampling. For a given input 2D spectral image, we perform $T$ stochastic forward passes through the network. Due to the active dropout layers in testing stage, each individual forward pass $t$ yields a predicted mean redshift, $\mu_t$, and a predicted aleatoric variance, $\sigma^2_t$. The final, combined redshift prediction, $\hat{\mu}$, is defined as the expectation value over all $T$ stochastic samples:
\begin{equation}\label{eq:mu}
    \hat{\mu} = \frac{1}{T}\sum_{t=1}^T\mu_t
\end{equation}
The total predicted variance, $\hat{\sigma}^2$, is subsequently formulated as the sum of two distinct terms: the mean aleatoric variance, which captures the inherent, data-driven noise of the observation, and the epistemic variance, which captures the model's inherent uncertainty as measured by the sample variance of the predicted means. This combined uncertainty is mathematically expressed as:
\begin{equation}\label{eq:sigma}
    \hat{\sigma}^2 = \frac{1}{T} \sum_{t=1}^{T} \sigma_t^2 + \frac{1}{T} \sum_{t=1}^{T} (\mu_t - \hat{\mu})^2
\end{equation}
By taking the square root of this total variance, we extract the final standard deviation, $\hat{\sigma}$, which represents the comprehensive uncertainty for each redshift measurement. 

\subsection{Training} \label{sec:training}
\begin{figure}
    \centering
    \includegraphics[width=\linewidth]{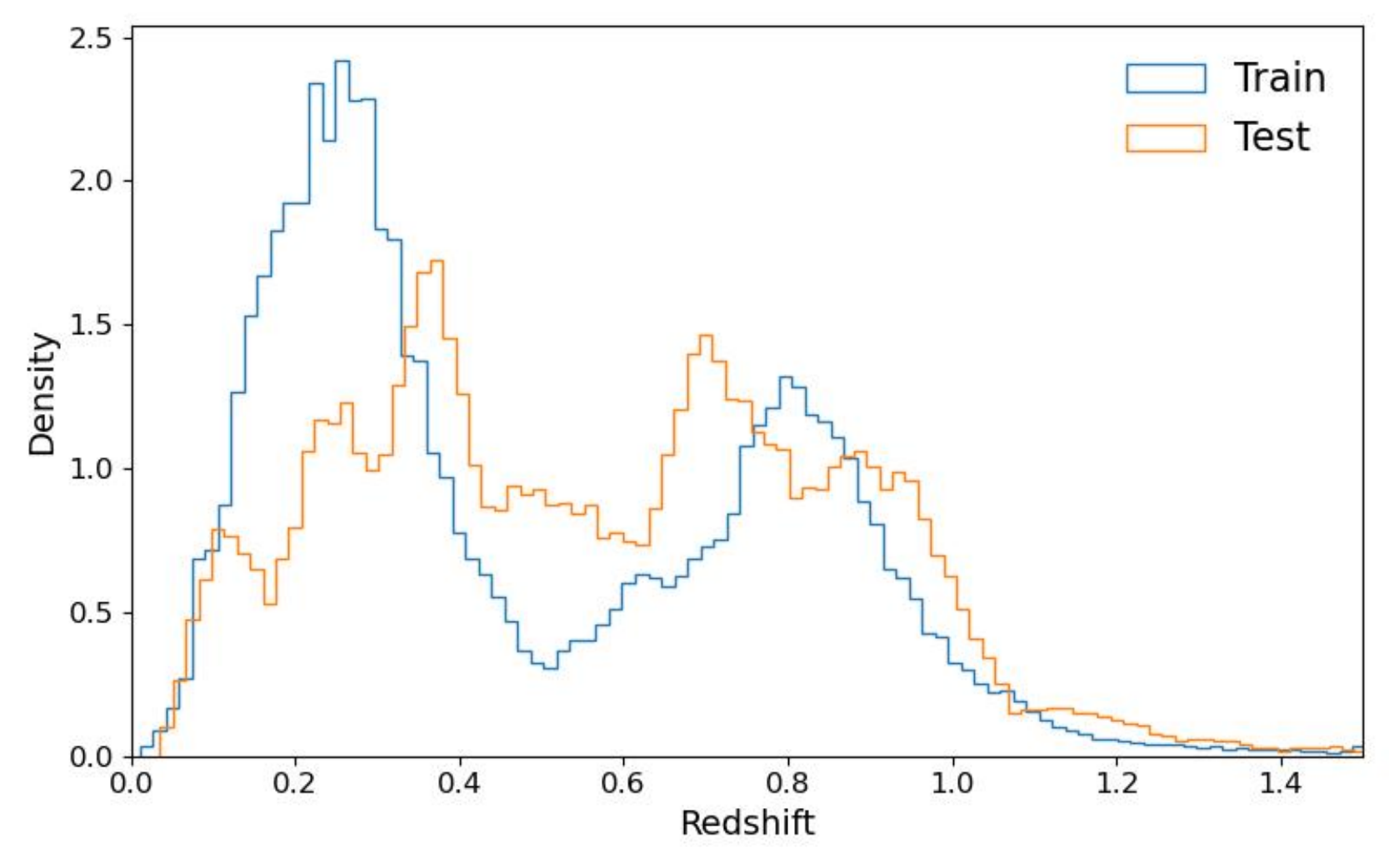}
    \caption{Redshift distributions of the training (blue) and testing (orange) datasets. The testing set is explicitly sampled to follow the anticipated distribution of CSST slitless spectroscopic survey.}
    \label{fig:data_split}
\end{figure}

The datasets described in Section~\ref{sec:datasets} are split into training and testing sets using an 8: 2 ratio. Rather than employing a random split, the testing set is constructed to mirror the anticipated spectroscopic redshift distribution of the upcoming CSST survey~\citep{Gong2019}. The resulting redshift distribution for both the training and testing sets are presented in Figure~\ref{fig:data_split}.

To stabilize the training of the Bayesian network, we adopt a transfer learning strategy~\citep{Pan2010,Zhuang2021,Farahani2021}. We initially train a deterministic version of the entire network to establish a robust feature space, and subsequently utilize the pre-trained feature extractor as the foundation for the Bayesian estimator, which yields the final redshift measurements alongside their corresponding uncertainties.

First, we pre-train the deterministic base network. This optimization is performed using a Mean Squared Error (MSE) loss function and the AdamW optimizer. AdamW improves upon the standard Adam algorithm by decoupling weight decay from the gradient updates, thereby providing more effective regularization~\citep{Loshchilov2017}. To ensure convergence of the model, we employ a cosine annealing learning rate schedule, gradually decaying the learning rate from an initial value of $1\times10^{-4}$ over maximum 100 epochs. To evaluate network performance, we primarily monitor normalized median absolute deviation (NMAD) $\sigma_{\rm NMAD}$, a precision metric that is robust against catastrophic outliers, defined as:
\begin{equation}\label{eq:sigma_nmad}
  \sigma_{\rm NMAD} = 1.48\times{\rm median}\left(\left|\frac{\Delta z - {\rm median}(\Delta z)}{1 + z_{\rm true}} \right|\right),
\end{equation}
where $\Delta z = z_{\rm pred} - z_{\rm true}$, with $z_{\rm pred}$ and $z_{\rm true}$ denoting the predicted and true redshifts, respectively. During training, we continuously monitor this metric, and save the model corresponding to the lowest $\sigma_{\rm NMAD}$ for testing stage. Although $\sigma_{\rm NMAD}$ serves as our primary criterion for model selection, it is unsuitable as a direct loss function for network optimization, since it is not differentiable at the center of the data, creating a flat gradient that makes it hard to optimize using gradient descent. Furthermore, while robust to outliers, the median operation does not utilize all data points to calculate the loss value, failing to provide enough information for efficient model convergence compared to the mean operation of MSE. 


To prevent overfitting and enhance the generalization of the network, we apply extensive data augmentation to the training set. This includes horizontal flipping and random translations along two spatial dimensions up to 10\% of the image pixels. Crucially, these spatial translations are explicitly designed to simulate the positional uncertainties inherent to slitless spectroscopy, which directly manifest as wavelength calibration errors. By applying one horizontal flip and six distinct random translations per image, we expand the effective size of our training set by a factor of eight. Note that we do not apply vertical flips or rotations, due to the rectangular geometry and the physically meaningful directional dependence of the spectral images. Through this augmentation strategy, the network learns to naturally account for spatial offsets, enabling it to robustly estimate redshifts from the 2D spectral images while entirely bypassing traditional wavelength calibration pipelines. 

For the Bayesian network, the training configuration remains largely consistent with the deterministic pre-training phase, but the objective function is replaced by the Negative Log-Likelihood (NLL) for a Gaussian distribution, defined as:
\begin{equation}\label{eq:nll}
  \text{NLL}(\mu_i, \sigma_i) = \sum_{i=1}^{n} \left[ \frac{1}{2} \log(2\pi\sigma_i^2) + \frac{(y_i - \mu_i)^2}{2\sigma_i^2} \right]
\end{equation}
where $\mu_i$ and $\sigma_i$ represent the network's predicted mean redshift and aleatoric standard deviation for the $i$-th sample, respectively, and $y_i$ is the true spectroscopic redshift. Notably, the standard deviation $\sigma_i$ is learned without direct supervision, since it can be solved by balancing the two terms in NLL. Additionally, the dropout rate is a hyperparameter and adjusted by experiments to 0.1. 

To identify the optimal Bayesian model, we continue to monitor $\sigma_{\rm NMAD}$ and retain the model weights corresponding to the lowest value. However, a single forward pass through the network yields only one stochastic sample from the predicted posterior, which may deviate from the true mean prediction. Therefore, to obtain a stable and representative estimate, we need to pass the testing data through the network multiple times and average the results. To balance computational efficiency with robust sampling during the validation steps of the training phase, we execute 10 MC forward passes for testing data and calculate the mean prediction and compute the $\sigma_{\rm NMAD}$. 

Raw uncertainties predicted by Bayesian neural networks are frequently mis-calibrated, often yielding systematically overconfident or underconfident bounds that do not accurately reflect the true error distribution. Statistically robust uncertainties need to be well-calibrated, meaning the empirical coverage probability of the predictions should precisely match the expected theoretical confidence interval (e.g., 68\% of true values fall within the predicted $1\sigma$ bounds). To correct for this, we apply temperature scaling, a robust post-training calibration technique. This method scales the predicted uncertainties by a globally optimized temperature parameter, explicitly aligning the network's output distribution with rigorous statistical principles. The deterministic pre-training and BCNN architectures are both implemented by \texttt{PyTorch}~\footnote{\url{https://pytorch.org/}}. The relevant scripts on data generation and training neural networks are publicly available in the Github~\footnote{\url{https://github.com/xczhou-astro/CSST_specimgs}}. 

\begin{figure}
    \centering
    \includegraphics[width=1\linewidth]{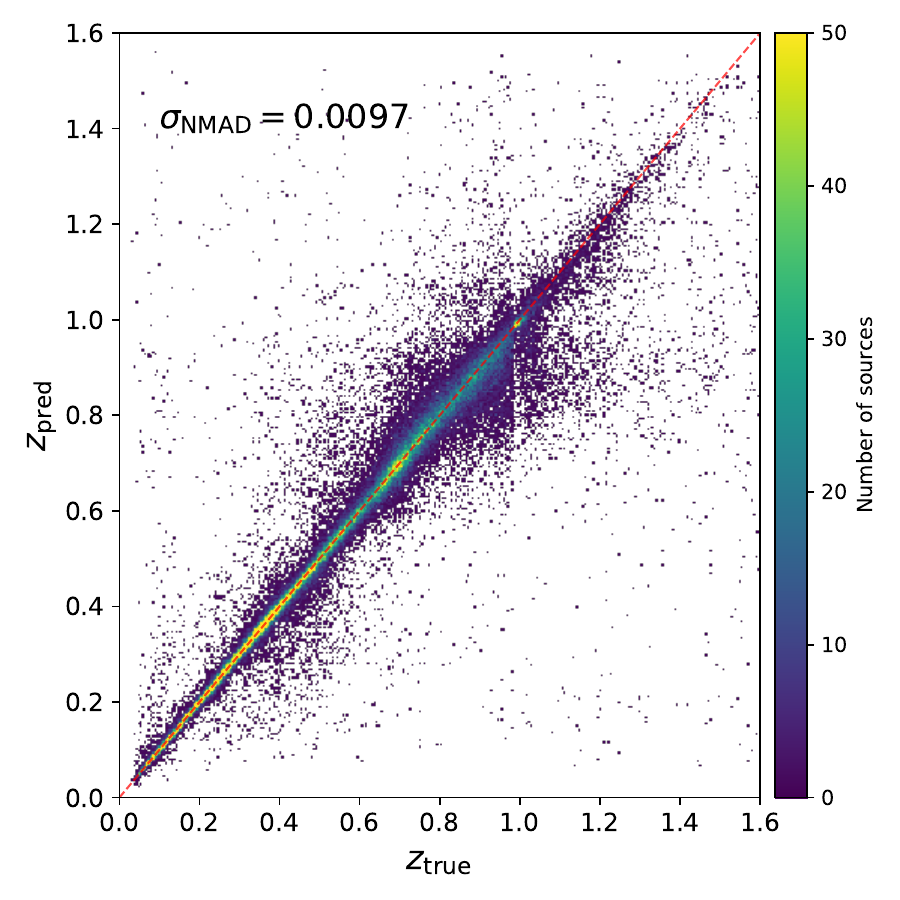}
    \caption{The density plot of the spec-$z$ predictions from the deterministic pre-training network. The color bar indicates the number of sources within each pixel. Overall, this base network achieves a precision $\sigma_{\rm NMAD}=0.0097$ for all sources with ${\rm SNR}_{GI}\geq1.0$. }
    \label{fig:point_result}
\end{figure}

\section{Results} \label{sec:results}


\begin{figure}
  \centering
  \includegraphics[width=1\linewidth]{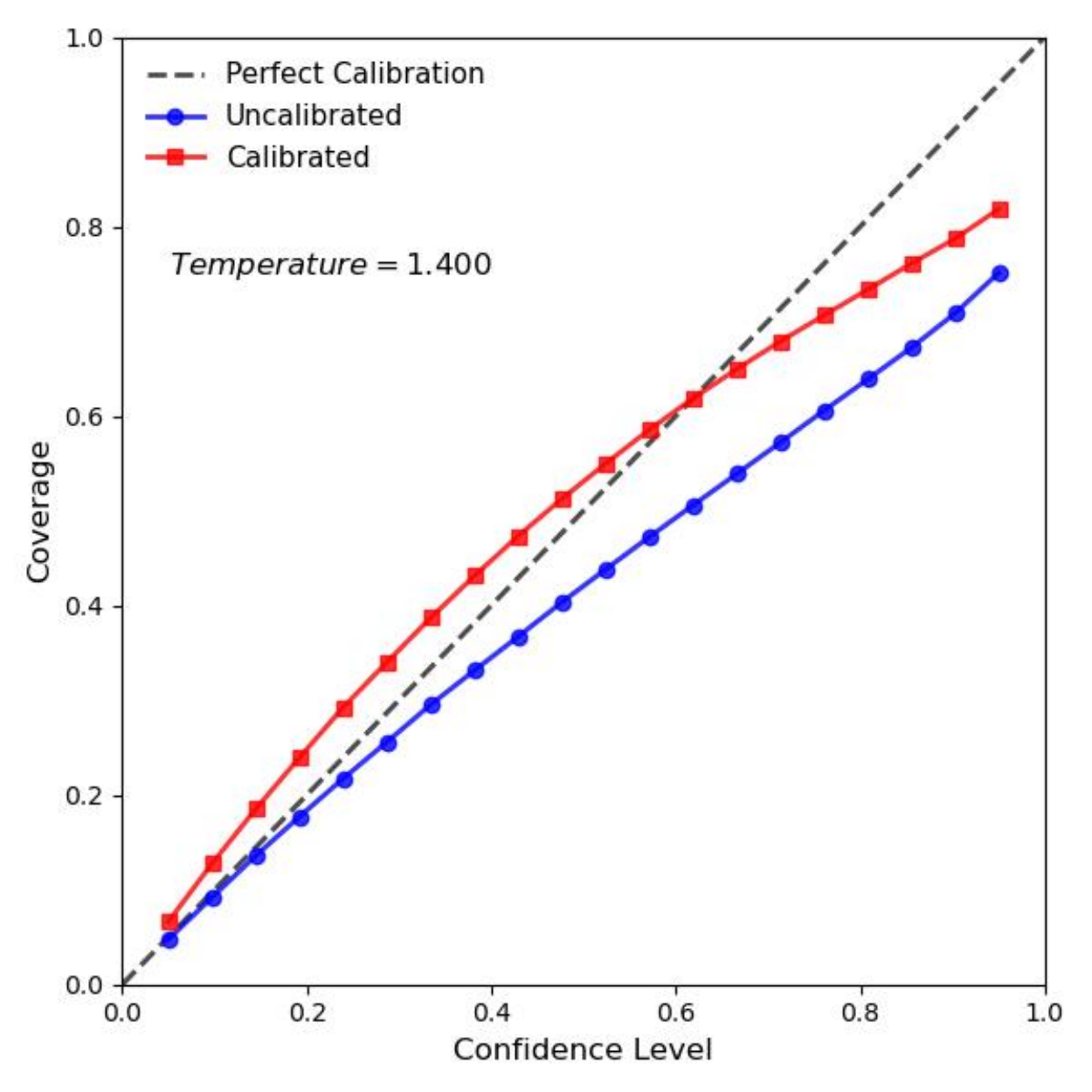}
  \caption{Confidence versus coverage plot used to evaluate the statistical reliability of the predicted uncertainties. The gray dashed, blue, and red curves illustrate the ideal, pre-calibration, and post-calibration scenarios, respectively. The derived temperature scaling factor of $T = 1.400$ indicates that the raw uncertainties were initially under-estimated.}
  \label{fig:qqplot}
\end{figure}

\begin{figure*}
    \centering
    \includegraphics[width=0.45\linewidth]{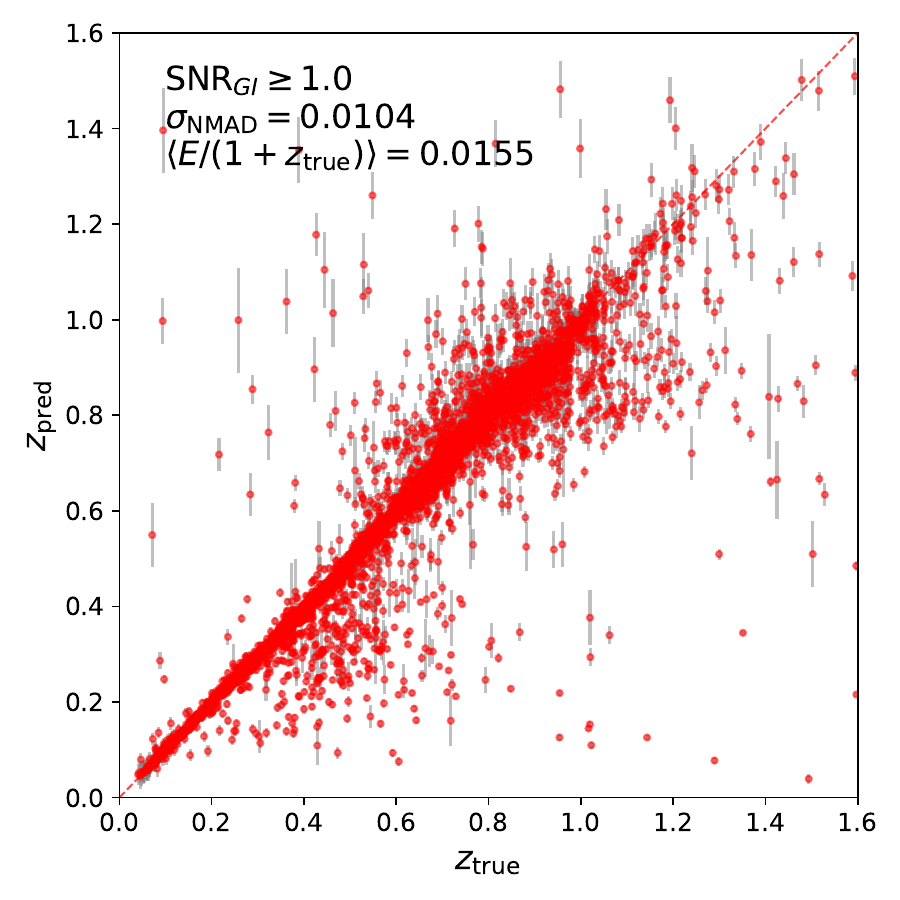}
    \includegraphics[width=0.45\linewidth]{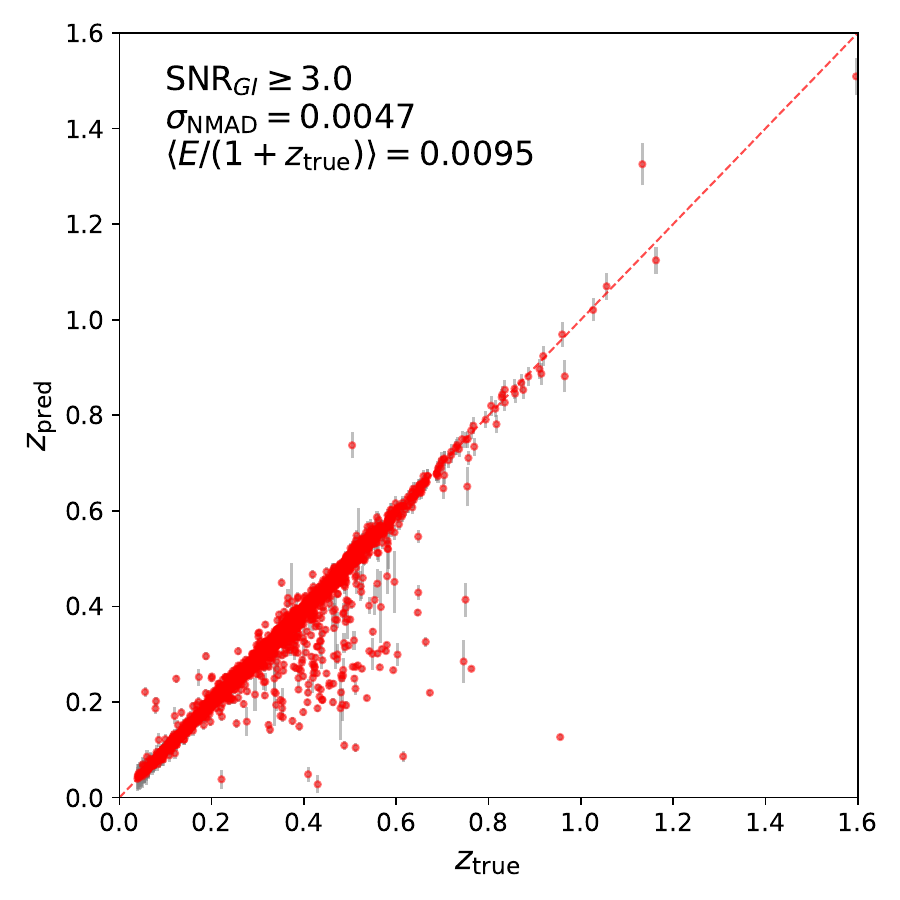}
    \\
    \includegraphics[width=0.45\linewidth]{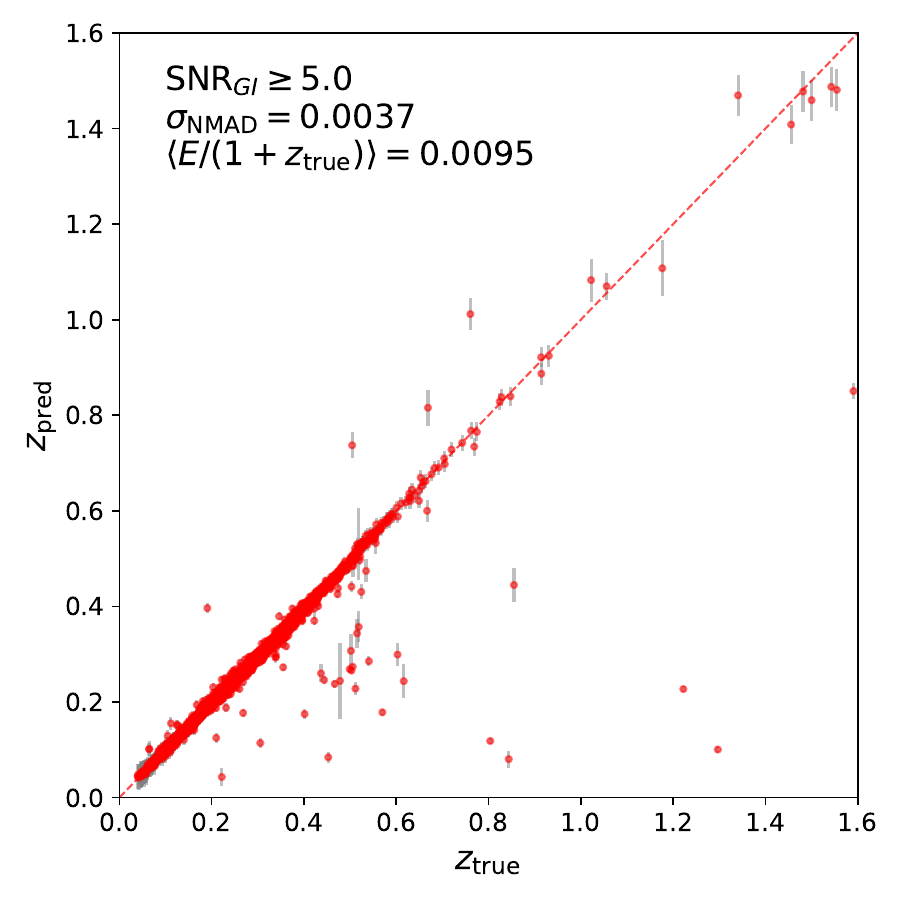}
    \includegraphics[width=0.45\linewidth]{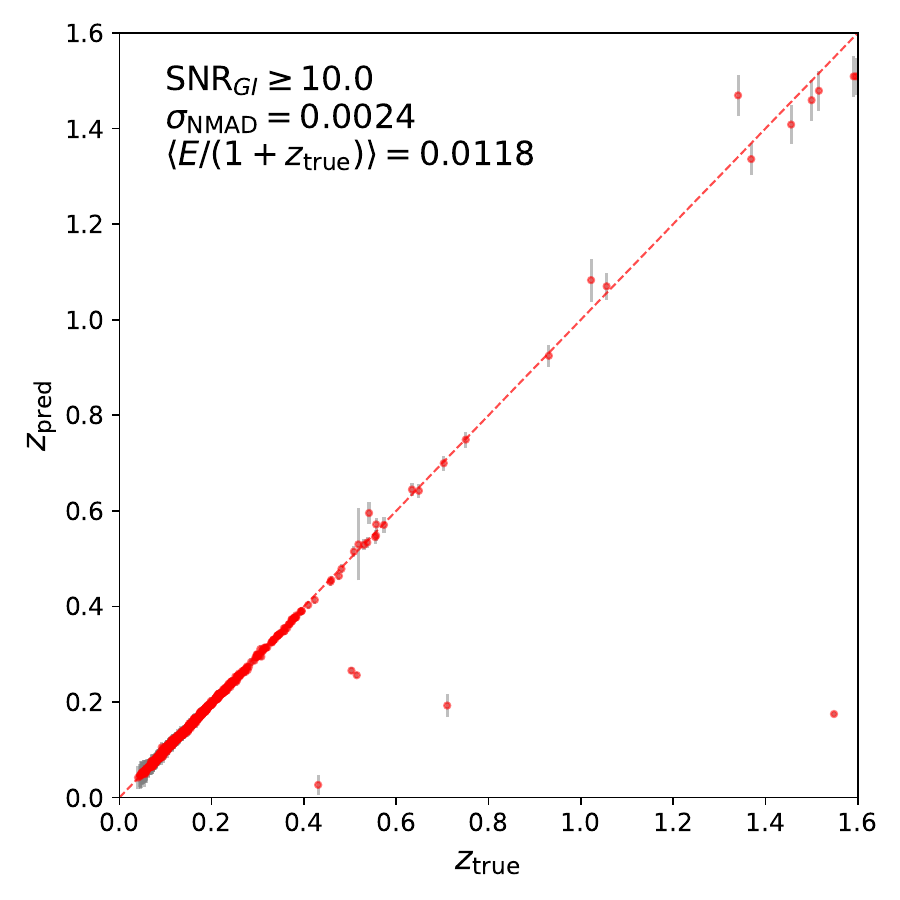}
    \caption{Predicted redshift $z_{\rm pred}$ versus true redshift $z_{\rm true}$ evaluated across four $GI$-band SNR thresholds ${\rm SNR}_{GI} \geq 1.0, 3.0, 5.0,$ and $10.0$. The error-bars indicate the calibrated total uncertainties. Each panel reports the corresponding predictive precision $\sigma_{\rm NMAD}$ and the mean normalized uncertainty $\langle E / (1 + z_{{\rm true}}) \rangle$. To maintain visual clarity, a random subsample of 5,000 sources is shown in the first three panels, while the final panel (${\rm SNR}_{GI} \geq 10$) displays the complete subset of 1,221 sources.}
    \label{fig:bayesian_result}
\end{figure*}

\begin{table}
  \caption{Performance metrics for the mock testing dataset, evaluated across various ${\rm SNR}_{GI}$ thresholds. The table reports the total number of sources $N$, the predicted precision $\sigma_{\rm NMAD}$, and the mean normalized uncertainty $\langle E / (1 + z_{{\rm true}}) \rangle$ for thresholds of ${\rm SNR}_{GI} \ge 1, 3, 5,$ and $10$.}
  \label{tab:result_stats}
  \centering
  \begin{tabular}{|l|l|l|l|l|l}
    \hline
    ${\rm SNR}_{GI}$            & $\geq$ 1  & $\geq$ 3   & $\geq$ 5 & $\geq$ 10  \\ \hline\hline
    $N$            & 86,196 & 28,900  & 10,630 & 1,221   \\ \hline
    $\sigma_{\rm NMAD}$ & 0.0104 & 0.0047 & 0.0037 & 0.0024 \\ \hline
    $\langle E / (1 + z_{{\rm true}}) \rangle$ & 0.0155 & 0.0095 & 0.0095 & 0.0118 \\ \hline
  \end{tabular}
\end{table}

\begin{figure}
  \centering
  \includegraphics[width=1\linewidth]{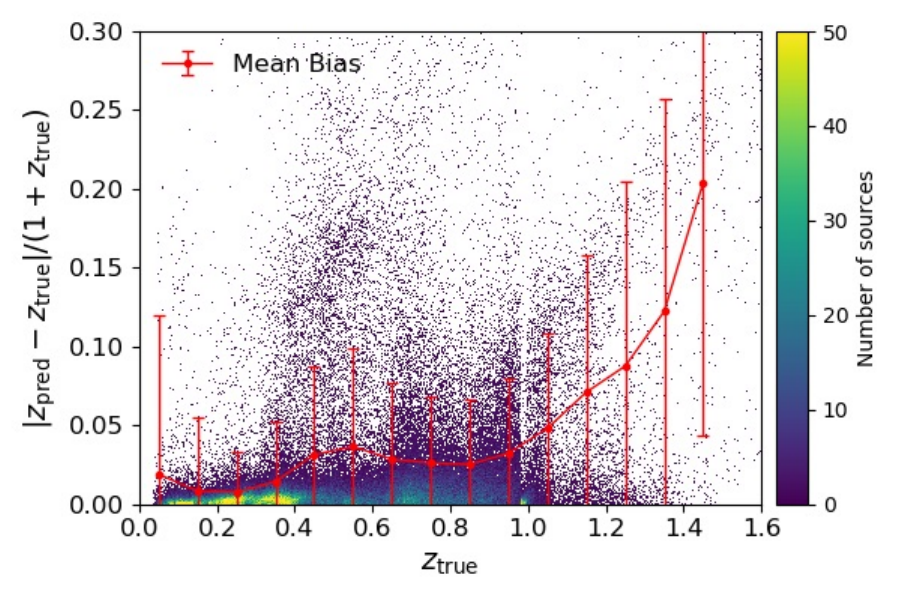}
  \\
  \includegraphics[width=1\linewidth]{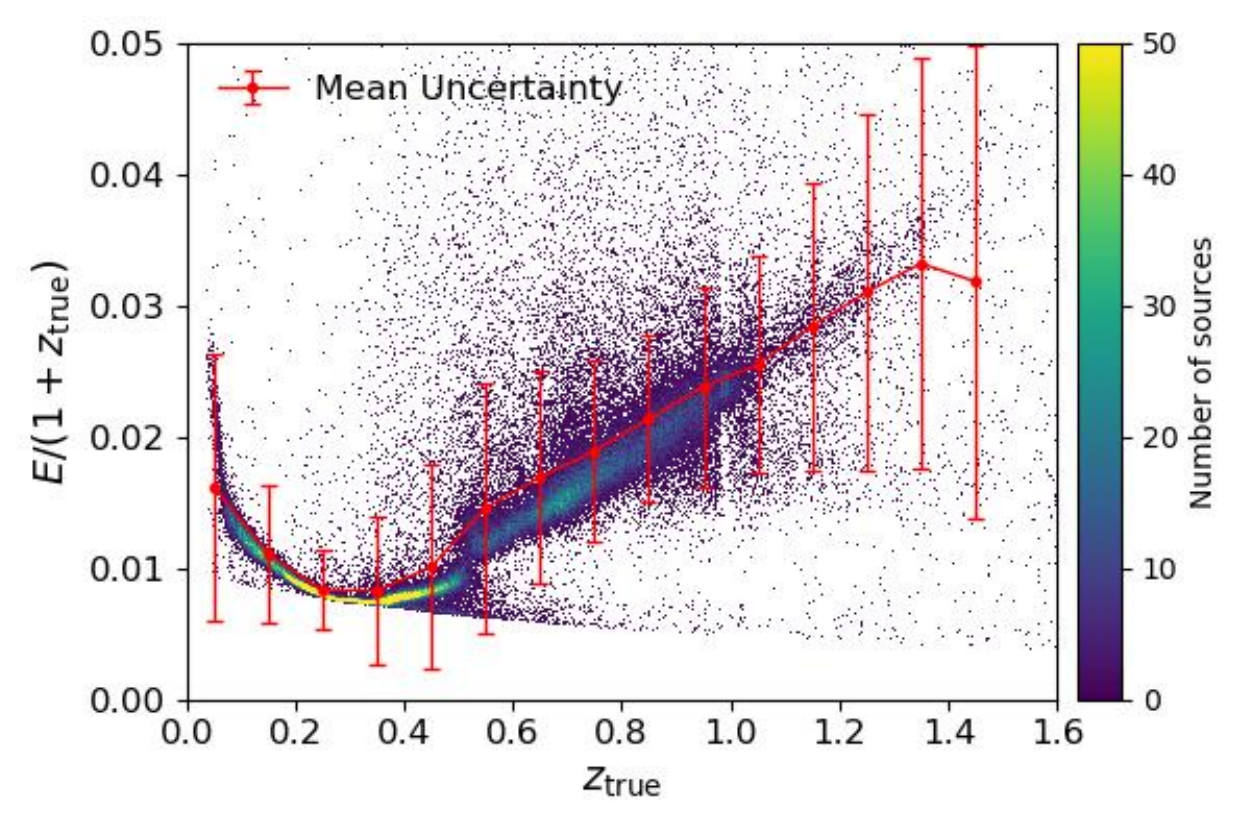}
  \caption{\textit{Upper}: The normalized absolute bias, $|z_{\rm pred} - z_{\rm true}|/(1 + z_{\rm true})$, as a function of true redshift $z_{\rm true}$. The color bar indicates the number of sources within each pixel. The red curve and corresponding error bars are the mean and standard deviation values calculated in redshift bins of $\Delta z = 0.1$. \textit{Lower}: The calibrated normalized uncertainty, $E / (1 + z_{{\rm true}})$, as a function of true redshift. The mean and standard deviation values are calculated using the same $\Delta z = 0.1$ binning as the upper panel.}
  \label{fig:stats_by_z_true}
\end{figure}

\begin{figure}
  \centering
  \includegraphics[width=1\linewidth]{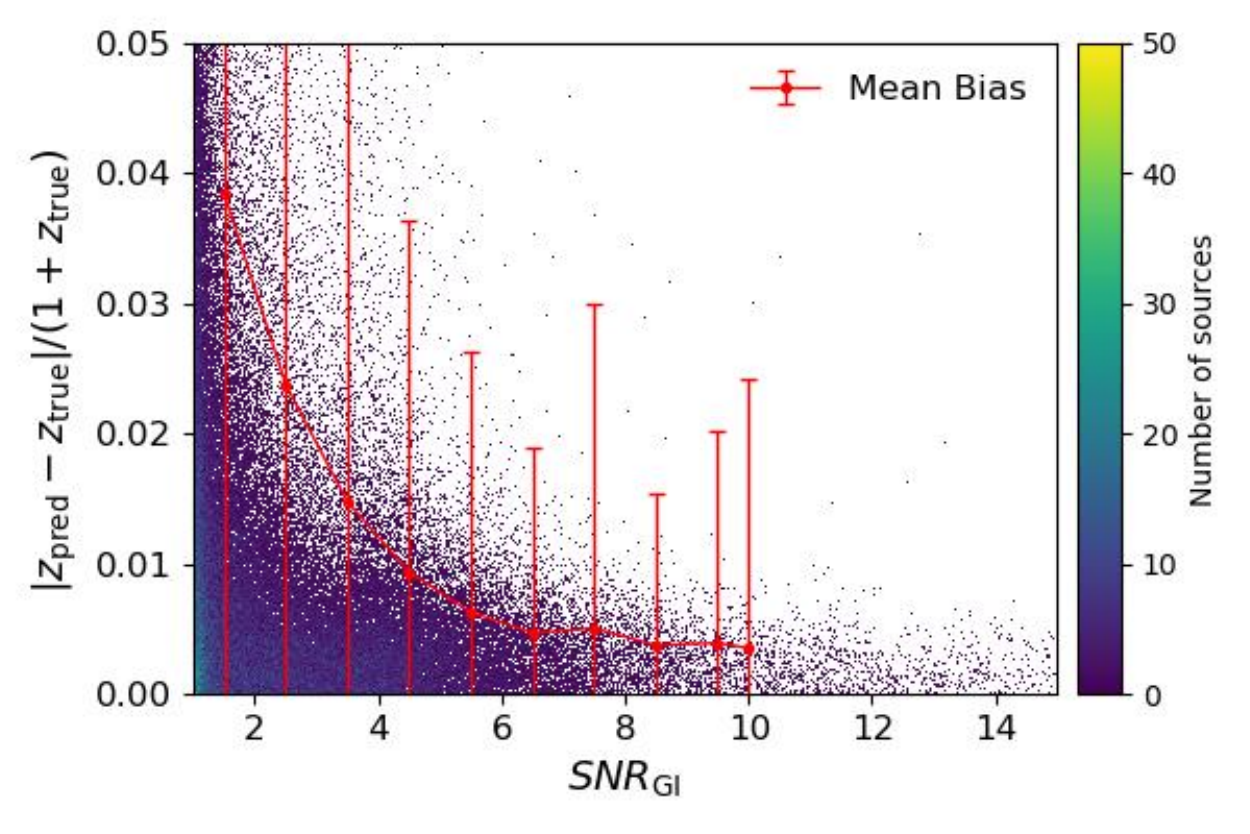}
  \\
  \includegraphics[width=1\linewidth]{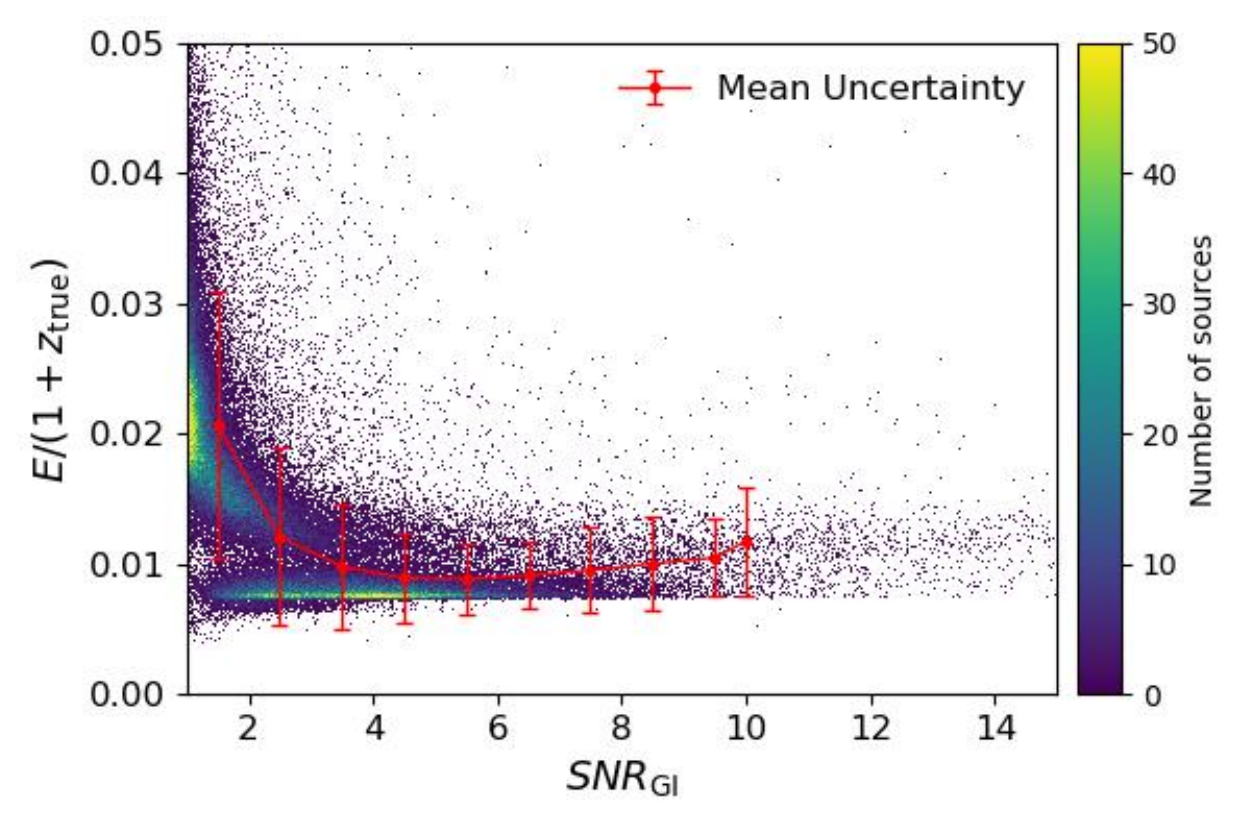}
  \caption{\textit{Upper}: The normalized absolute bias, $|z_{\rm pred} - z_{\rm true}|/(1 + z_{\rm true})$ as a function of ${\rm SNR}_{GI}$. The color bar indicates the number of sources within each pixel. The red curve and corresponding error bars are the mean and standard deviation values calculated in bins of $\Delta {\rm SNR}_{GI} = 1$. Note that the final data point at ${\rm SNR}_{GI} = 10$ are derived from all sources with ${\rm SNR}_{GI} \geq 10$. \textit{Lower}: The calibrated normalized uncertainty, $E / (1 + z_{{\rm true}})$, as a function of ${\rm SNR}_{GI}$. The mean and standard deviation values are evaluated using the identical $\Delta {\rm SNR}_{GI} = 1$ binning applied in the upper panel.}
  \label{fig:stats_by_snr}
\end{figure}

In addition to $\sigma_{\rm NMAD}$, we evaluate the performance of our framework using the normalized absolute bias, $|z_{\rm pred} - z_{\rm true}|/(1 + z_{\rm true})$ and the normalized uncertainty, $E/(1+z_{\rm true})$, where $E$ denotes the calibrated predicted uncertainty.

To establish a baseline for the predictive power of learning directly from 2D spectral images, we first evaluate the deterministic pre-trained network. Figure~\ref{fig:point_result} displays the predicted redshift $z_{\rm pred}$ against the true redshift $z_{\rm true}$ for the testing set. The deterministic model achieves a result $\sigma_{\rm NMAD}=0.0097$ for all sources with ${\rm SNR}_{GI}\geq1.0$. The concentration of data points along the diagonal line demonstrates that the CNN can effectively extract redshifts directly from the 2D images, successfully bypassing the need for explicit 1D spectral extraction and wavelength calibration. 

For the BCNN framework, we perform MC dropout sampling by feeding the testing data for 200 times to the network, and calculate the final mean point estimates and corresponding uncertainties using Equation~\ref{eq:mu} and Equation~\ref{eq:sigma}. As mentioned in Section~\ref{sec:training}, a fundamental requirement for Bayesian inference in precision cosmology is that the predicted uncertainties are statistically reliable. To evaluate this, we compute the empirical coverage probability of our posterior distributions. Figure~\ref{fig:qqplot} illustrates the coverage fraction against the theoretical confidence level. The uncalibrated BCNN output (blue line) falls below the perfect calibration diagonal. This indicates that the raw network is systematically overconfident, predicting lower uncertainty values. To correct this, we apply temperature scaling. We scale the uncertainties by a single global temperature $T=1.400$. The resulting calibrated coverage (red line) demonstrates better agreement with the theoretical expectation, ensuring that the uncertainties are statistically robust. 

Figure~\ref{fig:bayesian_result} displays the predicted redshift $z_{\rm pred}$ versus true redshift $z_{\rm true}$ evaluated across four $GI$-band SNR thresholds ${\rm SNR}_{GI}\geq$1.0, 3.0, 5.0 and 10.0. The error-bars indicate the calibrated total uncertainties. Each panel reports the corresponding predictive precision $\sigma_{\rm NMAD}$ and the mean normalized uncertainty $\langle E / (1+z_{\rm true}) \rangle$. To maintain visual clarity, a random subsample of 5,000 sources is shown in the first three panels, while the final panel (${\rm SNR}_{GI}\geq10.0$) displays the complete subset of 1,221 sources. For sources with ${\rm SNR}_{GI}\geq1.0$, the BCNN yields $\sigma_{\rm NMAD}=0.0104$. The slight degradation compared to the results of deterministic model shown in Figure~\ref{fig:point_result} is an expected trade-off in Bayesian learning, since the optimization shifts from minimizing the point estimate error to learning a distribution. Despite this, the BCNN maintains competitive accuracy while gaining the critical ability to quantify the uncertainties. Throughout the four panels, as expected, applying higher SNR cuts increases the concentration of data points along the diagonal line and enhances the precision metric. $\sigma_{\rm NMAD}$ improves continuously, resulting in 0.0104, 0.0047, 0.0037 and 0.0024 for sources with ${\rm SNR}_{GI}\geq$ 1.0, 3.0, 5.0 and 10.0. Particularly, the sources with ${\rm SNR}_{GI}\geq3.0$ meets the redshift requirements for studies such as BAO using CSST slitless spectroscopic surveys~\citep{Miao2024}. 

Concurrently, the mean normalized uncertainty $\langle E / (1+z_{\rm true}) \rangle$ initially decreases to 0.0155 to 0.0095 as the threshold is raised to ${\rm SNR}_{GI}\geq3$, reflecting a natural reduction in the data noise. Interestingly, for the high SNR cuts (${\rm SNR}_{GI}\geq10$), the mean uncertainty exhibits a slight, counter-intuitive increase. This behavior is primarily attributable to the insufficiency of data sample, where the statistic is significantly affected by the outliers. Table~\ref{tab:result_stats} summarizes the sample size $N$, precision $\sigma_{\rm NMAD}$, and the mean normalized uncertainty $\langle E / (1+z_{\rm true}) \rangle$ across different data quality regimes.

The upper panel of Figure~\ref{fig:stats_by_z_true} shows the normalized absolute bias $|z_{\rm pred} - z_{\rm true}|/(1+z_{\rm true})$ with respect to the true redshift. The red curve and its associated error-bars denote the mean and standard deviation of this metric, calculated within redshift bins of width $\Delta z=0.1$. Overall, the predicted performance exhibits a strong inverse correlation with the training source density (see Figure~\ref{fig:data_split}). At low redshifts ($z_{\rm true} < 0.4$), where training sources are abundant, the network achieves lowest bias. In the intermediate redshift ($z_{\rm true}\sim0.5$), a peak in the bias emerges, directly mirroring the drop in source density within this range. As we move to higher redshifts up to $z_{\rm true}\sim0.9$, the local density increase and the bias correspondingly decreases. Finally, beyond $z_{\rm true}\sim0.9$, the bias increases significantly as the availability of training sources continues to diminish. The lower panel displays the calibrated normalized uncertainty $E/(1+z_{\rm true})$ with respect to the redshifts. Unlike the mean absolute bias, which exhibits localized fluctuations across redshifts, the mean normalized uncertainty displays a continuous steady increase beyond $z_{\rm true}\sim0.4$. This behavior is physically and statistically expected: the continuous drop in SNR for high-redshift sources, as shown in Figure~\ref{fig:snr_vs_z}, causes the Bayesian network to appropriately output broader, less confident posterior distributions. Figure~\ref{fig:stats_by_snr} shows both the normalized absolute bias (upper panel) and the normalized uncertainty (lower panel) against ${\rm SNR}_{GI}$. As anticipated, both metrics decrease as the SNR increases, confirming that the BCNN correctly assigns higher confidence and achieves greater precision for observations with higher SNR.

\section{Discussions} \label{sec:discussions}

\begin{table}
\caption{The performance metrics for BGS, LRG and ELG sources according to the main target selection criteria of DESI. $N_{\rm total}$ and $N_{\rm testing}$ shows the total and testing data size, respectively.}
\label{tab:targets}
\centering
\begin{tabular}{|c|c|c|c|}
\hline
               & BGS     & LRG     & ELG    \\ \hline\hline
$N$            & 207,755 & 158,954 & 43,859 \\ \hline
$N_{\rm testing}$  & 31,912  & 39,900  & 10,790 \\ \hline
$\sigma_{\rm NMAD}$ & 0.0051  & 0.0177  & 0.0244 \\ \hline
$\langle E / (1 + z_{{\rm true}}) \rangle$            & 0.0093  & 0.0186  & 0.0246 \\ \hline
\end{tabular}
\end{table}

In this section, we evaluate the performance of our framework across different target classifications and present the results when simulated wavelength calibration errors are introduced. Additionally, we address several limitations of our current approach.

\subsection{Target classifications}\label{sec:target}

\begin{figure}
    \centering
    \includegraphics[width=1\linewidth]{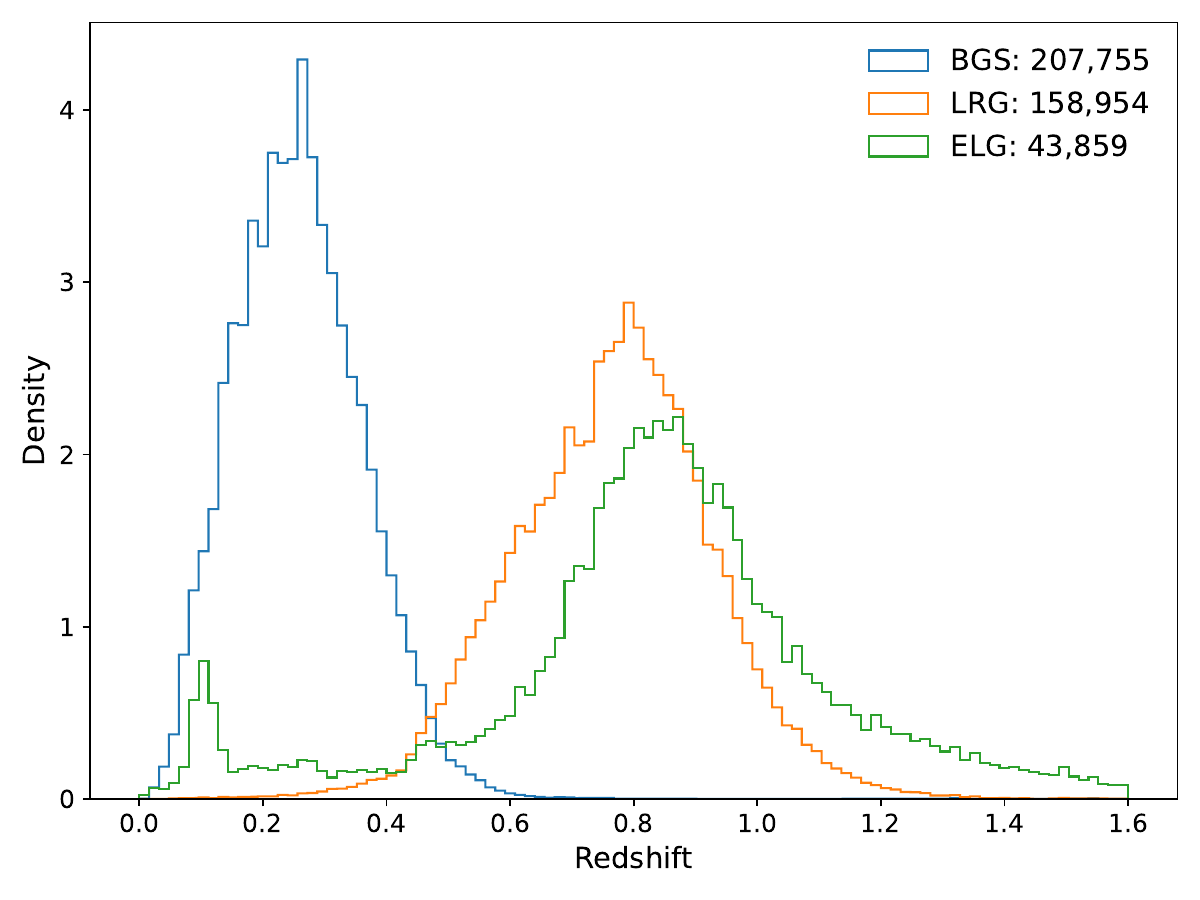}
    \caption{Redshift distributions for the BGS (orange), LRG (green), and ELG (blue) source populations, classified according to the main DESI target selection criteria. The total number of sources for each target class is provided in the legend.}
    \label{fig:desi_targets}
\end{figure}

Based on the main target selection criteria of the DESI survey, the galaxies can be categorized into three distinct classes, the Bright Galaxy Sample (BGS), Luminous Red Galaxies (LRG) and Emission Line Galaxies (ELG)~\citep{Hahn2023,Zhou2023target,Raichoor2023}. Table~\ref{tab:targets} presents the total sample sizes $N_{\rm total}$, the number of testing samples $N_{\rm testing}$, and the corresponding performance metrics for each of these classes. BGS constitutes the vast majority of our dataset, following by LRG, while ELG represent the rarest population. Their redshift distributions are illustrated in Figure~\ref{fig:desi_targets}. The BGS predominantly comprises bright and low-redshift sources with high SNR. In contrast, LRG trace intermediate-to-high redshifts, and ELG generally occupy the highest-redshift tail of the survey volume. We notice a decreased density at redshift $z\sim0.5$, consistent with the redshift distribution shown in Figure~\ref{fig:data_split}. 

Evaluating the performance across these distinct populations reveals a clear hierarchy. The BGS yield the highest precision $\sigma_{\rm NMAD}=0.0051$ alongside the lowest mean uncertainty $\langle E / (1 + z_{{\rm true}}) \rangle = 0.0093$. LRG exhibit intermediate performance, whereas ELG show the worst $\sigma_{\rm NMAD}=0.0244$ and highest uncertainty. This performance distinction is driven by a combination of physical and statistical factors. Physically, BGS targets provide high-SNR observations with prominent and easily identifiable spectral features across a wide optical continuum. Conversely, ELG are intrinsically fainter, and their redshift estimation relies heavily on narrow emission lines (such as the [OII] doublet). Because the resolution of slitless spectroscopy is low, these narrow lines often suffer from severe smearing, making them significantly harder for the network to resolve. On the other hand, the predicted accuracy correlate with the available sample size $N_{\rm total}$. The relative scarcity of ELG in training set inherently restricts the network's ability to thoroughly learn the mapping from the 2D spectral images to redshift values and their corresponding uncertainties. 

\subsection{Wavelength calibration}\label{sec:wavelength calibration}
To explicitly test our deep learning framework's resilience to the calibration errors, we simulate wavelength calibration failures by applying spatial perturbations to the testing dataset. Specifically, we introduce random spatial translations of up to 10\% of the image dimensions along both the horizontal and vertical axes. We evaluate these perturbed 2D spectral images using the trained Bayesian network for 200 runs as mentioned in Section~\ref{sec:results}.

As anticipated, introducing these artificial degrades the overall predicted precision. For subsets evaluated at ${\rm SNR}_{GI}$ threshold of $\geq1, 3, 5$ and $10$, the $\sigma_{\rm NMAD}$ becomes $0.0121,0.0056,0.0044$ and $0.0032$, respectively. Compared to the performance listed in Table~\ref{tab:result_stats}, this represents only a modest worse performance. This result clearly demonstrates that by operating directly on the 2D spectral images and leveraging the inherent translational invariance of convolutional networks, the deep learning framework can implicitly correct astrometric uncertainties. Consequently, this approach offers a highly robust alternative to bypass the complicated 1D wavelength calibration pipelines. 

\subsection{limitations}\label{sec:limitations}
While our BCNN demonstrates strong performance for redshift estimation directly from 2D slitless spectra, our current framework relies on several simplifying assumptions. To transition this method to real observational data from the CSST, future iterations must address the following limitations. 


In generating our mock 2D spectra, we utilized high-resolution HSC $i$-band images as the universal spatial template for each source. This approach assumes that the structural morphology of a galaxy remains constant across the entire wavelength range. In reality, galaxies exhibit significant color gradient. For example, star-forming regions are typically distributed in extended disks or distinct knots, which dominate the blue continuum, while older and redder stellar populations are concentrated in central bulges. Consequently, the true spatial profile is highly wavelength-dependent. Ignoring the wavelength dependent morphology may lead to biases in the 2D spectral images across different bands. 

Our current slitless spectroscopic simulation only considers the Poisson and Gaussian noises. However, true observations are subjected to a complex instrumental and environmental systematics, including cosmic ray, detector defects, complex background gradients from zodiacal light and spatially varying PSF on the focal plane. A deep learning model trained on idealized noises may struggle to generalize to the artifacts present in real data. To create more realistic datasets, future simulations need to incorporate a more comprehensive forward-modeling pipeline that mimics the precise instrumental signature and background complexities of the CSST detectors. 

\section{Conclusion}\label{sec:conclusion}
In this work, we present a deep learning framework designed to estimate spec-$z$ directly from 2D slitless spectral images, entirely bypassing the complicated 1D spectrum extraction. To train and validate our approach, we constructed a highly realistic CSST mock dataset of over 680,000 spectra in $GV$ and $GI$ bands, from high-resolution galaxy images from HSC-SSP PDR3 and SEDs from DESI DR1. 

By employing a BCNN implemented by Monte Carlo dropout, our deep learning model can accurately estimate spec-$z$ values with their corresponding uncertainties accounting for both epistemic and aleatoric aspects from 2D spectral images. The deterministic baseline network achieves a precision of $\sigma_{\rm NMAD}=0.0097$ for sources with ${\rm SNR}_{GI}\geq1$, while the BCNN maintains a high precision $\sigma_{\rm NMAD}=0.0104$ by transferring the feature extractor of the deterministic baseline model. By applying higher SNR threshold to be 3.0, 5.0 and 10.0, $\sigma_{\rm NMAD}$ can reach 0.0047, 0.0037 and 0.0024 respectively, matching the redshift accuracy requirements for studies such as BAO employing CSST slitless spectroscopic survey. Following a temperature scaling calibration, the BCNN yields reliable predicted uncertainties, resulting in mean normalized uncertainties $\langle E / (1 + z_{{\rm true}}) \rangle$ of 0.0155, 0.0095, 0.0095 and 0.0118 respectively for ${\rm SNR}_{GI}\geq1.0, 3.0, 5.0$ and $10.0$ respectively. These uncertainties successfully reflect physical and observational constraints, appropriately scaling inversely with the SNR. 

Evaluating the network across standard survey populations reveals an expected performance hierarchy. The BGS yields the highest precision, followed by LRG, while the fainter, higher-redshift ELG exhibits largest scatter due to their smeared narrow emission lines and smaller training size. 

To mimic wavelength calibration failure, we simulate spatial translations of up to 10\% of the image dimensions. In this case, the network experiences a modest degradation in performance, achieving $\sigma_{\rm NMAD}=0.0121, 0.0056, 0.0044$ and $0.0032$ for sources with ${\rm SNR}_{GI}\geq1.0,3.0,5.0$ and $10.0$. This robustness proves its viability as a compelling alternative to traditional data analysis pipelines that rely on extracted 1D spectra. 

Ultimately, our framework establishes a powerful proof-of-concept for the direct analysis of 2D slitless spectra. Deploying robust and uncertainty-aware deep learning models directly on 2D spectral images offers a highly promising pathway to unlock the cosmological potential of the CSST and other next-generation slitless spectroscopic missions. 

\begin{acknowledgments}
  X.Z. and Y.G. acknowledge the support from the CAS Project for Young Scientists in Basic Research (No. YSBR-92), National Key R\&D Program of China grant Nos. 2022YFF0503404 and 2020SKA0110402. This work is also supported by science research grants from the China Manned Space Project with grant Nos. CMS-CSST-2025-A02, CMS-CSST-2021-B01, and CMS-CSST-2021-A01. HM acknowledges support from the National Natural Science Foundation of China (NSFC, grant No. 12503008). NL acknowledge the support from the CAS Project for Young Scientists in Basic Research (No. YSBR-062). 
\end{acknowledgments}





%
\facilities{CSST, HSC, DESI}




\bibliography{sample701}{}
\bibliographystyle{aasjournalv7}



\end{document}